\documentclass[showpacs,amsmath,amssymb, nobibnotes, aps, prl,showkeys]{revtex4}
\usepackage{graphicx}
\usepackage{dcolumn}
\usepackage{bm}
\usepackage{docs}
\usepackage{amssymb}
\usepackage{amsmath}
\usepackage{bigints}
\usepackage{relsize}
\expandafter\ifx\csname package@font\endcsname\relax\else
 \expandafter\expandafter
 \expandafter\usepackage
 \expandafter\expandafter
 \expandafter{\csname package@font\endcsname}%
\fi

\newcommand{\ltsima} {$\; \buildrel < \over \sim \;$}
\newcommand{\gtsima} {$\; \buildrel > \over \sim \;$}
\newcommand{\lta} {\lower.5ex\hbox{\ltsima}}
\newcommand{\gta} {\lower.5ex\hbox{\gtsima}}

\newcommand{\RNum}[1]{\uppercase\expandafter{\romannumeral #1\relax}}

\begin{document}

\title[Gravitational time delay of particle with non-zero mass]{Effects of dark energy and flat rotation curve on the gravitational time delay of particle with non-zero mass}


\author {Tamal Sarkar$^{1}$, Shubhrangshu Ghosh$^{2}$ and Arunava Bhadra$^{1}$  }

\affiliation{$^{1}$ High Energy $\&$ Cosmic Ray Research Centre University of North Bengal, Post N.B.U, Siliguri 734013, India.  }
\affiliation{$^{2}$ Centre for Astroparticle Physics and Space Science, Department of Physics, Bose Institute, Block EN, Sector V, Salt Lake, Kolkata, India 700091 }

\begin{abstract}

The effects of several dark energy models on gravitational time delay of particles with non-zero mass are investigated and analytical expressions for the same are obtained at the first order accuracy. Also the expression for gravitational time delay under the influence of conformal gravity potential that well describes the flat rotation curve of spiral galaxies is derived. The findings suggest that i) the conformal gravity description of dark matter reduces the net time delay in contrast to the effect of normal dark matter and therefore in principle the models can be discriminated using gravitational time delay observations and ii)the effect of dark energy/flat rotation curve may be revealed from high precision measurements of gravitational time delay of particles involving megaparsec and beyond distance-scale.     

\end{abstract}

\pacs{98.62.Mw, 04.70.Bw, 95.30.Sf, 04.20.-q, 04.20.DW}
\keywords{Gravitational time delay, dark energy, flat rotation curve, particle}

\maketitle


\section{Introduction}

The explanation of the observed accelerated expansion of the Universe requires that the bulk of the energy density in the Universe is repulsive, which is termed as dark energy (DE). On the other hand, galactic rotation curves and few other observations demand for existence of a non-luminous matter component, dubbed dark matter (DM). Several independent analysis of astrophysical and cosmological data now firmly suggest that DE, DM and luminous matter constitute about $68\%$, $27\%$ and $5\%$ of the total energy budget of the universe [1]. 

Out of the several wishful candidates for DE, the simplest candidate is the cosmological constant ($\Lambda$). The model involving cosmological constant, the so called $\Lambda$CDM model with a value of $\Lambda$ nearly $10^{-56} {\rm {cm}^{-2}}$ and CDM refers to cold DM, provides an excellent fit to the wealth of high-precision observational data, on the basis of a remarkably small number of cosmological parameters [2,3,4]. But the physical origin of cosmological constant remains a major problem. Besides due to its non evolving nature, the $\Lambda$CDM model suffers from the so-called coincidence problem [5]. Alternative candidates of DE include scalar-field models like quintessence [6,4], k-essence [7,8] and phantom field [9]. There are also proposals for modification of general relativity to account for the accelerated expansion without the need for DE which include scalar-tensor theories (or equivalently f(R) theories) [10,11], conformal gravity [12], massive gravity theories [13] including Dvali-Gabadadze-Porrati (DGP) braneworld gravity [14], etc. The nature of DM is also unknown at present but DM is a testable proposition in direct-detection experiments unlike DE. The gravitational effect of DM within a galaxy depends on the radial density profile of DM. While the flat rotation curve feature implies a logarithmic gravitational potential, the rotation curve data points for a large sample of spiral galaxies were also found to describe well by a gravitational potential linear in $r$ [15].     

Both DE and DM are expected to influence gravitational phenomena at all distance scales including those in the solar system. In solar system the influence of DE has been studied mainly through cosmological constant and is found to be maximum in the case of perihelion shift of mercury orbit where the $\Lambda$ contribution is about $10^{-15}$ of the total shift [16] and measurements of advances in the perihelia of Mercury imposes an upper limit $\Lambda < 10^{-42} m^{-2}$ [17]. On the other hand, analysis of the perihelion precession of Mercury, Earth, and Mars give the upper bound on the density of DM $\rho_{\rm dm} < 3 \times 10^{-19}$ $g/{\rm cm^{3}}$ [18]. Note that  the rotation curve data suggests that the density of DM in the Milky Way at the location of solar system is $\rho_{\rm dm} = 0.5 \times 10^{-24}$ $g/{\rm cm^{3}}$ [19]. DE is mainly effective at cosmological (megaparsec) scales and as a result the contribution of DE could be significant (larger than the second order term) even in a local gravitational phenomenon when kiloparsec (Kpc) to megaparsec (Mpc) scale distances are involved, such as the gravitational bending of light by cluster of galaxies [20] or the relativistic accretion phenomena around massive BHs (see [21] and references therein; [22]), whereas the effect of DM is significant at the outer part of galaxies. Consequently, for large distance scales, astrophysical and cosmological phenomena are likely to be dictated by DM/DE and hence to probe DE/DM from local phenomena, one has to explore the local gravitational phenomena involving Kpc to Mpc distance-scale.  

The phenomenon of gravitational time delay of an electrically neutral (henceforth termed as just neutral throughout the rest of the manuscript) particle with non-zero mass such as neutrino/neutron from an extra-galactic source may offer a possibility of studying the influence of DE/DM as it involves Mpc distance-scale. Here it is worthwhile to mention that, measurements of gravitational time delay, for example, an extra time delay that light suffers while propagating in gravitational field over the time required for light transmission between two points in Minkowski spacetime, through Doppler tracking of the Cassini spacecraft on its way to the Saturn, currently imposes the most stringent constraint on the first parameterized post-Newtonian parameter $\gamma$ with $\gamma -1 < (2.1\pm2.3) \times 10^{-5}$ [23]. Note that $\gamma$ is zero in the Newtonian theory, unity in general relativity, and $\gamma -1$ is considered as a measure of a deviation from general relativity. However, the effect of cosmological constant on gravitational time delay of photon is comparatively less and solar system measurements give only the restriction $\Lambda \le 10^{-24} m^{-2}$ [16]. The effect of DE on gravitational time delay of photons has already been investigated by Asada [24]. Very recently, the effect of DE/DM on gravitational time advancement (negative effective time delay) has been investigated by Ghosh and Bhadra [25]. Considering a neutral particle for time delay measurement (as well as other similar effects) is advantageous over photon due to the fact that the time delay for particle depends also on the mass and energy of the particle, thereby offering additional control on the measurement [26] and we shall argue later in the discussion section that this additional control should be useful to study experimentally gravitational time delay involving Kpc distance scale.  

In the present work we derive the analytical expression of gravitational time delay for particles having non-zero rest mass considering the presence of DE and DM and discuss about the experimental feasibility to test DE/DM effects on time delay experimentally in future. The letter is organized as follows: in the next section we would formulate the problem mathematically for gravitational time delay corresponding to a neutral particle with non-zero rest mass for general spherically symmetric static spacetime. In \S 3 we derive the analytical expression for time delay in presence of DE/DM considering up to first order in $M/r$ and $\beta$, where $M$ is the mass of the gravitating object, $\beta$ is the parameter describing the strength of DE/DM. In \S 4 we discuss our results stressing the possibilities of experimental detection of such effects in a future experiment. Finally, we conclude in \S 5.

\section{Gravitational time delay of a neutral particle with mass}

For a general static spherically symmetric metric of the form

\begin{eqnarray}
ds^2= c^{2}{\mathcal{B}}(r) dt^2 -{\mathcal{A}}(r)dr^2 - r^2 d\Omega^2
\label{1}
\end{eqnarray}

where, $d\Omega^2 = d\theta^2 + \sin^2 \theta d\phi^2 $ and $c$ the usual speed of light, the geodesic equations for a test particle motion in equatorial plane around a spherical matter distribution having mass $M$ lead to the following relation 

\begin{eqnarray}
\frac{{\mathcal{A}}(r)}{{{\mathcal{B}}(r)}^{2}}\left(\frac{dr}{dt}\right)^{2}+\frac{\alpha_1}{r^2}-\frac{c^2}{{\mathcal{B}}(r)} = -\alpha_2 c^2\, , 
\label{2}
\end{eqnarray}

where $\alpha_1$ is the specific relative angular momentum $\left[\alpha_1 \equiv r^{4} \left(\frac{d\phi}{dp}\right)^2 \right]$, $\phi$ is the azimuthal angle, $p$ is the affine parameter describing the trajectory) which is a constant of motion and $\alpha_2$ is a normalization constant connecting proper time $\tau$ and $t$ (${d\tau}^{2}= \alpha_{2} \,{dp}^{2}$). For particle with non-zero rest mass $m$, $\alpha_2 > 0$; whereas for particle with zero rest mass, $\alpha_2 = 0$. At the distance of closest approach $r_p$, $\frac{dr}{dt}$ must vanish, i.e., $\frac{dr}{dt} {\vert}_{r=r_p}= 0 $. This gives
\begin{eqnarray}
\alpha_1 = c^2 \left[- \alpha_2 + \frac{1}{{\mathcal{B}}(r_p)}\right]r_{p}^{2} \, .
\label{3}
\end{eqnarray}

From Eqs. (2) and (3), we obtain the time required by a particle to traverse a distance from $r_p$ to $r$, which is given by

\begin{eqnarray}
t \, (r,r_p)=\frac{1}{c} \int_{r_p}^{r}\sqrt{{\mathcal{P}}(r,\alpha_2)} \, dr \, , 
\label{4}
\end{eqnarray}

where, 

\begin{eqnarray}
{\mathcal{P}}(r,\alpha_2) \approx \frac{{\mathcal{A}}(r)/{\mathcal{B}}(r)}{\left[1-\alpha_2 \, 
{\mathcal{B}}(r) + \frac{{r_p}^{2}}{r^{2}}\left(\alpha_2 \,{\mathcal{B}}(r) 
- \frac{{\mathcal{B}}(r)}{{\mathcal{B}} (r_p)} \right ) \right]} \, .
\label{5}
\end{eqnarray}

Once the spacetime geometry is given, the gravitational time delay can be computed from the Eq. (4) through Eq. (5).  
Restricting up to first order in $M$, where $M$ is mass of the gravitating object, the total travel time in the Schwarzschild geometry (${\mathcal{B}}(r)={\mathcal{A}}(r)^{-1}=1-2GM/c^{2}r$),  is given by [27]

\begin{eqnarray}
t_{\rm {Sch}} \left(r,r_p \right)  \approx \frac{1}{c \sqrt{1-\alpha_{2}}} \left\{ \sqrt{r^{2}-r_{p}^{2}} +  \frac{GM\left(2-3 \alpha_{2}\right)}{c^{2}\left(1-\alpha_{2}\right)} \ln\frac{\left(r+\sqrt{r^{2}-r_{p}^{2}}\right)}{r_{p}} + \frac{GM}{c^{2}\left(1-\alpha_{2}\right)}\sqrt{\frac{r-r_{p}}{r+r_{p}}} 
     \right\} \, ,
\label{6}
\end{eqnarray} 

where, subscript `Sch' represents `Schwarzschild'. When $\alpha_{2}=0$, i.e. for photons, the above equation reduces to the well known expression for gravitational time delay of photons. 
    
\section{Gravitational time delay of a neutral particle with mass in presence of DE/DM}

In presence of DE/DM the exterior vacuum spacetime will be no longer Schwarzschild geometry but a modified one. Here we shall consider the following form of the metric tensor

\begin{equation}
{\mathcal{B}}(r)=1-\frac{2Gm}{c^2 r}-\beta_{1} r^{n}
\label{7}
\end{equation}
and
\begin{equation}
{\mathcal{A}}(r)=1+\frac{2Gm}{c^2 r}+\beta_{2} r^{n} \, , 
\label{8}
\end{equation}

where, $\beta_{1}$ and $\beta_{2}$ are constants. Different choices of n, $\beta_{1}$ and $\beta_{2}$ lead to different models of DE/DM. 

case 1: With $n=1/2$, $\beta_{1}=2\beta_{2}=\pm 2\sqrt{GM/r_{c}^{2}}$ the model represents the gravitational field of a spherically symmetric matter distribution on the background of an accelerating universe in DGP braneworld gravity provided leading terms are only considered [28]. $r_{c}$ is the crossover scale beyond which gravity becomes five dimensional.  

case 2: For the choice $n=1$, with $\beta_{1}=\beta_{2} = -\beta = -\left(5.42 \times 10^{-42} \frac{M}{M_{\odot}} + 3.06 \times 10^{-30}\right)$ ${\rm cm}^{-1}$ the model well describes the flat rotation curves of spiral galaxies [12,15]. 

case 3: If $n=3/2$, and $\frac{2}{3}\beta_{1}=-\beta_{2}=m_{g}^{2}\sqrt{\frac{2GM}{13c^{2}}}$, the model corresponds to the non-perturbative solution of a massive gravity theory (an alternative description of accelerating expansion of the universe) [29] where $m_{g}$ is the mass of graviton. 

case 4: When $n=2$, $\beta_{1}=\beta_{2}=\Lambda/3$ and $m=\mu$ the above metric describes the Schwarzschild-de Sitter (SDS) or Kotler space-time which is the exterior space time due to a static spherically symmetric mass distribution in presence of the cosmological constant 
$\Lambda$ with $\Lambda \sim10^{-56} {\rm {cm}^{-2}}$ [30].  

Note that $m$ and $M$ have been defined in \S 2. Using DE/DM led by ${\mathcal{A}}(r)$ and ${\mathcal{B}}(r)$, restricting up to first order correction due to $\beta_{i}$ (i=1,2), and neglecting the terms of the order $M^2$ and beyond, ${\mathcal{P}}(r,\alpha_2)$ given by 
Eq. (5) reduces to the form, given by 

\begin{eqnarray}
\sqrt{{\mathcal{P}}(r,\alpha_2)} \approx \frac{1   + \frac{GM\left(2-3 \alpha_{2}\right)}{c^{2}r\left(1-\alpha_{2}\right)}   + \frac{GMr_{p}}{c^{2}\left(1-\alpha_{2}\right)r(r+r_p)} +
\left(\frac{\beta_1+\beta_2}{2} - \frac{\beta_1 \alpha_2}{2 \left(1-\alpha_2 \right)}  \right) \, r^n  - \frac{\beta_1 r^2_p \, \left(r^n - r^n_p \right)}{2 \left(1-\alpha_{2}\right) \left(r^2 - r^2_p \right)} }{\sqrt{\left(1-\frac{r_{p}^{2}}{r^{2}}\right)\left(1- \alpha_{2}\right)}} \, .
\label{9}
\end{eqnarray}

Using Eqs. (4) and (9), we obtain the explicit expression to compute the time required by a particle to traverse a distance from $r_p$ to $r$ in the presence of spacetime geometry defined by Eq. (1) with Eqs. (7) and (8), corresponding to different DE/DM models, given by 

\begin{eqnarray}
t_{n} \left(r,r_p \right)  \approx t_{\rm {Sch}} \left(r,r_p \right) + \frac{1}{2 \, c \, \sqrt{1-\alpha_{2}}} \left\{ 
      \left[\beta_1 + \beta_2 - \frac{\beta_1 \alpha_2}{\left(1-\alpha_2 \right)} \right] \, {\mathcal{I}}^1_n  
     - \frac{\beta_1 }{ \left(1-\alpha_2 \right)} \, {\mathcal{I}}^2_n   \right\} \, ,
\label{10}
\end{eqnarray} 
\\
where, ${\mathcal{I}}^1_n$ and ${\mathcal{I}}^2_n$ are integrals defined by ${\mathcal{I}}^1_n = \bigintss_{r_p}^r \frac{r^{n+1} \, dr}{\sqrt{\left(r^2-r^2_p \right)}}$ and ${\mathcal{I}}^2_n = r_{p}^{2} \bigintss_{r_p}^r \frac{r \left(r^n - r^n_p \right) \, dr}{\left(r^2-r^2_p \right) \sqrt{\left(r^2-r^2_p \right)}}$. (\ref{10}) is the general expression for the time required to traverse a distance from $r_p$ to $r$ in the presence of a generic metric given by Eqs. 
(1), (7) and (8) corresponding to DE/DM models to first order corrections. For $n=1$ and $n=2$ corresponding to DM and DE respectively, we have analytical 
solutions of ${\mathcal{I}}^1_1$, ${\mathcal{I}}^2_1$ and ${\mathcal{I}}^1_2$, ${\mathcal{I}}^2_2$ which are given below

\begin{eqnarray}
{\mathcal{I}}^1_1 = \frac{r}{2} \sqrt{r^2-r_{p}^2} +  \frac{r_{p}^{2}}{2}\, \ln \frac{r+ \sqrt{r^2-r_{p}^2} }{r_p} , \nonumber \\
{\mathcal{I}}^2_1 = -r_{p}^{2}\sqrt{\frac{r-r_p}{r+r_p}} + r_{p}^{2} \ln \frac{r+ \sqrt{r^2-r_{p}^2} }{r_p}  \, .
\label{11}
\end{eqnarray}

\begin{eqnarray}
{\mathcal{I}}^1_2 = \frac{1}{3} \sqrt{r^2-r_{p}^2} \left(r^2+2r_{p}^{2} \right) \, , \nonumber \\
{\mathcal{I}}^2_2 = r_{p}^{2} \sqrt{r^2-r_{p}^2} \, .
\label{12}
\end{eqnarray}

For general $n$, however, ${\mathcal{I}}^1_n$ and ${\mathcal{I}}^2_n$ can only be expressed through hypergeometric functions which is not very useful. Under the assumption $r >> r_p$, the integrals ${\mathcal{I}}^1_n$ and ${\mathcal{I}}^2_n$ for general $n$ ($n \ne 1$ ) reduce to 

\begin{eqnarray}
{\mathcal{I}}^1_n \approx \frac{r^{n+1}}{n+1} \, + \, \frac{r_{p}^{2} \, r^{n-1}}{2(n-1)} \, , \nonumber \\
{\mathcal{I}}^2_n \approx r_{p}^{2} \frac{r^{n-1}}{n-1} \, . 
\label{13}
\end{eqnarray}

For $\alpha_2=0$, Eq. (10) with Eq. (13) gives the gravitational time delay for photon which agrees with the results obtained by Asada [24] at the leading order in r. In deriving the contribution of $\beta_{i}$ on gravitational time delay we ignore the cross terms between M and $\beta_{i}$ (and higher order terms in $\beta_{i}$) since $\beta_{i}$ is small. However, in some circumstances the cross terms may be relevant which are given in the appendix. 
 
\section{Discussions}

The Eq. (10) through Eqs. (11)-(13) imply that DE enhances the time delay effect. Similar effects of DE were noted earlier for photons [16,24]. For gravitational time delay the influences of DE is somewhat counter intuitive; the repulsive nature of DE is expected to act differently than normal mass. Note that the gravitational bending angle of photon is reduced by the repulsive nature of DE [31]. The potential linear in radial distance that describes the flat rotation curves of spiral galaxies well, reduces the net time delay which is opposite to the effect of normal dark matter on gravitational time delay and hence the conformal gravity description of galactic rotation curve can be discriminated from normal dark matter from gravitational time delay measurement involving Mpc distance scale, at least in principle.   

The Eq. (10) with Eqs. (11) and (12) suggest that the DE and DM contribution to the time delay effect will be of the same order to the pure Schwarzschild contribution at distance scale roughly 30 Kpc and 300 Kpc respectively in our galaxy. Hence to detect the influences of DE/DM through gravitational time delay effect one needs to conduct the measurements involving Kpc distance-scale. Experimentally gravitational time delay is studied in solar system by measuring the round-trip travel time of an electromagnetic signal emitted from the Earth past the Sun to a planet or satellite and returned back to the Earth. Such a strategy is of course impractical for measuring gravitational time delay when Kpc scale distance is involved. Instead a feasible approach is to study the time difference of arrival of neutral particles with same mass at two (or more) different energies from a stellar collapse scenario such as extragalactic gamma ray bursts (GRBs) or supernova explosions (SNe). 

The theories of stellar collapse [32,33] demand that neutrinos of different energies should emit in a short pulse of about $10$ msec duration which is also indicated experimentally [34,35]. The photons are expected to emit a few hours later than the neutrino emission [32,33]. Since gravitational time delay by the galaxy causes a time delay about $5$ months [36], so the difference in arrival times of neutrinos and photons from extragalactic GRBs or SNe also may probe DE/DM influence on gravitational time delay. We, therefore, shall evaluate analytical expressions for difference in arrival times between neutral particles having non-zero rest mass (such as neutrinos) of two different energies and that between particle and photon.     

We consider the scenario that the particle/photon is emitted from the source $S(R)$, where `$R$' represents the radial coordinate of source `$S(R)$', reaches the observer $O(r_o)$ at $r_o$ (here the earth) with the distance of closest approach $r_p$. All the distances are measured taking the center of the spherical mass distribution with mass $M$ as the center of the coordinate system. Hence the total transit time is $T_n \left(r_o, R \right) = t_n \left(r_o,r_p \right) + t_n \left(r_p, R \right) \, $. The particle/photon emitted from the source suffers a gravitational time delay due to the spherical mass distribution. For the test particle with mass $m$ and energy $\varepsilon$ as received by the observer, the parameter $\alpha_2$ (as described in Eq. (2) ) and other subsequent equations in \S 2, \S 3) is given by [26]; $\alpha_2 = \frac{m^2}{{\mathcal{B}}(r_o) \varepsilon^{2}} = \frac{m^2}{\varepsilon^{2}} \left(1+\frac{2GM}{c^2 \, r_o}  + {\beta_1 r^n_o} \right)$ (restricting up to first order correction due to $\beta_1$).  
As we wish to focus on highly relativistic particles, we consider $\varepsilon >> m$. Denoting $\Delta T_{n} \vert^{\varepsilon_2}_{\varepsilon_1}$ and $\Delta T_{n} \vert^{m=0}_{m}$ as the difference in arrival times between particles with same mass $m$ but two different energies $\varepsilon_1$ and $\varepsilon_2$ and the difference in arrival times between particle with mass $m$ and energy $\varepsilon$ and photon, respectively, and restricting to the first order in the expansion of $M$ and $m^2/\varepsilon^2$, we get the most general expressions (ignoring the cross terms between $M$ and $\beta_{i}$): 

For the DM model with $n=1$ and $\beta_1=\beta_2 = -\beta$, 

\begin{align}
\Delta T_{1} \vert^{\varepsilon_2}_{\varepsilon_1} & \approx \frac{D_S \, m^{2}}{2 \, c} \left( \frac{1}{\varepsilon^2_1} - \frac{1}{\varepsilon^2_2} \right) \left\{1+  \frac{2GM}{c^2} \left[\frac{1}{r_{o}} + \frac{3}{2 D_S} \left(\sqrt{\frac{R -r_p}{R+r_p}} + \sqrt{\frac{r_o -r_p}{r_o+r_p}}  \right)  \right]    \right. \nonumber \\
    &\mathrel{\phantom{=}} \left.\kern-\nulldelimiterspace - \,  \beta \, \left[ r_o + \frac{3}{2} \frac{r^2_p}{D_S} \left(\sqrt{\frac{R -r_p}{R+r_p}} + \sqrt{\frac{r_o -r_p}{r_o+r_p}}  \right)  \,- \, \frac{3}{2} \frac{r^2_p}{D_S} \ln \frac{\left(R +  \sqrt{R^2-r^2_p} \right) \left( r_o  + \sqrt{r^2_o -r^2_p}  \right)}{r^2_p}  \right] \right\} \, , 
\label{14}
\end{align}
\\
where, $D_S = \sqrt{R^2-r^2_p} + \sqrt{r^2_o -r^2_p}$, the distance of the source $S(R)$ from the observer $O(r_o)$ on earth.

\begin{align}
\Delta T_{1} \vert^{m=0}_{m} & \approx \frac{D_S \, m^{2}}{2 \, c \, \varepsilon^{2}}  \left\{1+  \frac{2GM}{c^2} \left[ \frac{1}{r_{o}} + \frac{3}{2 D_S} \left(\sqrt{\frac{R -r_p}{R+r_p}} + \sqrt{\frac{r_o -r_p}{r_o+r_p}}  \right)  \right]    \right. \nonumber \\
    &\mathrel{\phantom{=}} \left.\kern-\nulldelimiterspace - \,  \beta \, \left[ r_o + \frac{3}{2} \frac{r^2_p}{D_S} \left(\sqrt{\frac{R -r_p}{R+r_p}} + \sqrt{\frac{r_o -r_p}{r_o+r_p}}  \right)  \,- \, \frac{3}{2} \frac{r^2_p}{D_S} \ln \frac{\left(R +  \sqrt{R^2-r^2_p} \right) \left( r_o  + \sqrt{r^2_o -r^2_p}  \right)}{r^2_p}  \right]  \right\} \, , 
\label{15}
\end{align}

whereas, corresponding to DE (described by cosmological constant), with $n=2$ and $\beta_1=\beta_2= \frac{\Lambda}{3}$ in SDS geometry, 

\begin{eqnarray}
\Delta T_{2} \vert^{\varepsilon_2}_{\varepsilon_1} \approx  \frac{D_S \, m^{2}}{2 \, c} \left( \frac{1}{\varepsilon^2_1} - \frac{1}{\varepsilon^2_2} \right) \left\{1+  \frac{2GM}{c^2} \left[ \frac{1}{r_{o}} + \frac{3}{2 D_S} \left(\sqrt{\frac{R -r_p}{R+r_p}} + \sqrt{\frac{r_o -r_p}{r_o+r_p}}  \right)   \right]   
     + \,  \lambda \, \left(\frac{r^2_o}{3} - \frac{r^2_p}{2}  \right) \right\} \, , 
\label{16}
\end{eqnarray}

\begin{eqnarray}
\Delta T_{2} \vert^{m=0}_m \approx  \frac{D_S \, m^{2}}{2 \, c\, \varepsilon^{2}}  \left\{1+  \frac{2GM}{c^2} \left[\frac{1}{r_{o}} + \frac{3}{2 D_S} \left(\sqrt{\frac{R -r_p}{R+r_p}} + \sqrt{\frac{r_o -r_p}{r_o+r_p}}  \right)  \right]   
     + \,  \lambda \, \left(\frac{r^2_o}{3} - \frac{r^2_p}{2}  \right) \right\} \, , 
\label{17}
\end{eqnarray}

and for general $n$, with condition $R >> r_p$; equivalently, $D_S \sim R, r_p \sim r_o$, we have

\begin{align}
\Delta T_{n} \vert^{\varepsilon_2}_{\varepsilon_1} &\approx  \frac{D_S m^{2}}{2 \, c} \left( \frac{1}{\varepsilon^2_1} - \frac{1}{\varepsilon^2_2} \right) \left\{1+  \frac{2GM}{c^2} \frac{1}{r_{o}} + \frac{(\beta_2 - \beta_1) R^{n}}{2} \left[ \frac{1}{n+1}+\frac{r_p^2}{2(n-1)R^2} \right]  + \beta_1 \left[r^n_o -  
\frac{3 \, r^2_o \, R^{n-2}}{2 (n-1)}\right] \right\} \, , 
\label{18}
\end{align}
 
\begin{align}
\Delta T_{n} \vert^{m=0}_m &\approx  \frac{D_S \, m^{2}}{2 \, c\, \varepsilon^{2}} \left\{1+  \frac{2GM}{c^2} \frac{1}{r_{o}} + \frac{(\beta_2 - \beta_1) R^{n}}{2} \left[ \frac{1}{n+1}+\frac{r_p^2}{2(n-1)R^2} \right]  + \beta_1 \left[r^n_o -  
\frac{3 \,r^2_o \, R^{n-2}}{2 (n-1)}\right] \right\} \, , 
\label{19}
\end{align}
respectively.

To find out the DM/DE contribution to gravitational time delay that the signal (like neutrino with mass `$m$') suffers while travelling from a distant source to the observer on earth about gravitating mass distribution with mass $M$, it is necessary to estimate in actual seconds, the quantities $\Delta T_{1} \vert^{m=0}_m$ or $\Delta T_{1} \vert^{\varepsilon_2}_{\varepsilon_1} $, $\Delta T_{2} \vert^{m=0}_m$ or $\Delta T_{2} \vert^{\varepsilon_2}_{\varepsilon_1} $, explicitly for terms corresponding to DM and DE, respectively. To compute these terms explicitly due to contribution of DM/DE as well as to clearly reveal the effects of DM and DE we focus on two (mathematically) simple but practically feasible scenarios as described hereunder. 

In one scenario, the signal originates from a distant-source, and the observer $O(r_o)$ is located  on the earth at a distance $r_o$ from the center of the spherical mass distribution. This can be represented by the condition $R >> r_p$ and $r_p \sim r_o$; which imply $D_S >> r_p$, $D_S >> r_o$. Such a scenario will arise, for example, if a signal originates from a distance local extragalactic source due to supernova explosion like that in the case of SN 1987A and suffers gravitational time delay by our galaxy. Another example of such a scenario is that when the source is an extragalactic one, situated far away from our local group and the signal from the source suffers gravitational time delay due to the contribution of the local group itself while reaching to the observer on the earth. 

Corresponding to this scenario, it can be seen from the Eqs. (16) and (17) that the difference in arrival time between particles with same mass but different energies $\varepsilon_1$ and $\varepsilon_2$ with $\varepsilon_2>\varepsilon_1$ or between particle with mass $m$ and energy $\varepsilon$ and photon is reduced due to cosmological constant; for the flat rotation curve similar aspect is also noticed from Eqs. (14) and (15). To ascertain the effective contribution to gravitational time delay due to DM/DE as compared to the pure Schwarzschild contribution for the cases described by the Eqs. (14) to (17), it is necessary to compute the magnitude of the ratio $(\chi)$ between these two contributing terms given in Eqs. (14) to (17). For the scenario described here, for the DM model, this ratio $\chi \approx \left\lvert \frac{-\beta c^2 r^2_o}{2 \, GM} \right\rvert$. Similarly, for the DE model (described by $\lambda$), $\chi \approx \left\lvert \frac{-\lambda c^2 r^3_o}{12 \, GM} \right\rvert$. The differences in time of arrivals would then be \\
for DM model:

\begin{equation}
\Delta T_{1} \vert^{m=0}_m \approx \left\lvert \frac{- \beta r_o D_S m^2 }{2 \,c \,  \varepsilon^2} \right\rvert 
\end{equation}
\begin{equation}
\Delta T_{1} \vert^{\varepsilon_2}_{\varepsilon_1} \approx \left\lvert \frac{- \beta r_o D_S m^2 }{2 \, c} \left(\frac{1}{\varepsilon^2_1} -\frac{1}{\varepsilon^2_2} \right)  \right\rvert , 
\end{equation}						
and for DE model:					
\begin{equation}						
\Delta T_{2} \vert^{m=0}_m \approx \left\lvert \frac{\lambda r^2_o D_S m^2 }{12 \, c \, \varepsilon^2} \right\rvert
\end{equation}	
\begin{equation}					
\Delta T_{2} \vert^{\varepsilon_2}_{\varepsilon_1} \approx \left\lvert \frac{\lambda r^2_o D_S m^2 }{12 \, c} \left(\frac{1}{\varepsilon^2_1} -\frac{1}{\varepsilon^2_2} \right)  \right\rvert. 
\end{equation}	

In the other scenario, the source $S(R)$ may be situated near to the center of the spherical mass distribution, however, $S(R)$ is located far away from the observer on the earth. This can be represented by the condition $R \sim r_p$; which imply $D_S \sim r_o$, $r_o >> r_p$. Such a scenario will arise when a core-collapse extragalactic SNe/GRB occurs near to the center of our local group and the signal originating from SNe/GRB is detected by the observer on the earth. For the scenario described 
here, $\chi \approx \left\lvert \frac{- \beta c^2 r^2_o}{5 \, GM} \right\rvert$ for the DM model, whereas for the DE model (described by $\lambda$), $\chi \approx \left\lvert \frac{\lambda c^2 r^3_o}{15 \, GM} \right\rvert$. 

For this scenario, corresponding to DM model describing the galactic rotation curves, the expressions of difference in arrival times between particles with same mass but different energies $\varepsilon_1$ and $\varepsilon_2$ with $\varepsilon_2>\varepsilon_1$ or between particle with mass $m$ and energy $\varepsilon$ and photon are identical to those in earlier mentioned scenario as given by Eqs. (20) to (21) whereas for cosmological constant model describing DE, the  differences in arrival times are just double to those in the previous scenario as given in Eqs. (22) and (23). Note that the galactic rotation curve feature reduces the net time delay, whereas for the cosmological constant, the reverse (enhance) aspect is noticed from Eqs. (16) and (17), which should be a distinguishing signature between DE and DM. 

In Table 1, we display the numerical estimate of the quantity $\chi$ for few potential astrophysical events (different signal sources and different gravitational mass distributions $(G.M)$ about which the signal suffers gravitational time delay), corresponding to the two scenarios described here, for both DM and DE (described by $\lambda$) models. It is to be noted that corresponding to the two scenarios described here, the ratio of the contributing terms to the gravitational time delay $(\chi)$ for both DM and DE (described by $\lambda$) models is independent of the distance of source from the observer, as well as of the closest distance of approach, however, only depends on the distance of the observer on earth from the gravitational mass distribution about which the signal suffers time delay. In Table 1 we choose appropriate gravitating systems $(G.M)$ and values of mass of gravitating systems about which the signal (like neutrino signal) suffers gravitational time delay, and corresponding values of $r_o$.

\begin{table}
\large
\centerline{\large Table 1}
\centerline{\large Values of $\chi$ for potential astrophysical events}
\begin{center}
\begin{tabular}{cccccccccccc}
\hline
\hline
\noalign{\vskip 2mm}
${\rm Scenarios}$ & ${\rm G.M}$ & $M/M_{\odot}$ & $r_o $ & $\chi$  & $\chi$  \\
$ $ & $ $ & $ $ & $ $ & $(\rm DM) $ & $(\rm DE)$ \\
\hline
\hline
\noalign{\vskip 2mm}
$R >> r_p$  &  ${\rm Our \, galaxy}$ & $\sim 10^{11}$  & $\sim 10 \, {\rm Kpc}$  & $\sim 10^{-1} $ & $\sim 10^{-6}$\\
\noalign{\vskip 2mm}
$ $  &  ${\rm Local \, group}$  & $\sim 10^{12}$ & $\sim  0.52 \, {\rm Mpc}$ & $\sim 7 $ & $\sim 2\times 10^{-2}$\\
\noalign{\vskip 2mm}
$ $  &  ${\rm Local \, supercluster}$  & $\sim 10^{15}$ & $\sim  16 \, {\rm Mpc}$ & $\sim 7000 $ & $\sim 0.6$\\
\hline
\noalign{\vskip 2mm}
$R \sim r_p$  & ${\rm Our \, galaxy}$  & $\sim 10^{11}$  & $\sim 10 \, {\rm Kpc}$   & $\sim 4 \times 10^{-2} $ & $\sim 8 \times 10^{-7}$  \\
\noalign{\vskip 2mm}
$ $  &  ${\rm Local \,  group}$  & $\sim 10^{12}$ & $\sim  0.52 \, {\rm Mpc}$   & $\sim 2.8 $  & $\sim 1.6 \times 10^{-2}$\\
\noalign{\vskip 2mm}
$ $  &  ${\rm Local \, supercluster}$  & $\sim 10^{15}$ & $\sim  16 \, {\rm Mpc}$ & $\sim 2800 $ & $\sim 0.5$\\
\hline
\hline
\end{tabular}
\end{center}
\end{table}

In Table 2, we estimate the values of differences in arrival times for similar choice of gravitating systems as in Table 1. For the case $R >> r_p$, corresponding to our local group, we choose a distant core-collapse SNe emitting neutrinos, located at a typical distance of $\sim 10$ Mpc from the observer on earth whereas for local supercluster we choose the source at $50$ Mpc away. Similarly for the case $R \sim r_p$, corresponding to our local group, we choose a core-collapse SNe emitting neutrinos from near to the center of the mass of the local group; for which $D_S \sim r_o$. Corresponding to our galaxy about which the signal suffers gravitational time delay, for the case $R >> r_p$, we choose a SNe neutrino emitting source to be located at a typical distance of $\sim 50$ Kpc from the observer on earth; whereas, for the case $R \sim r_p$, we choose a SNe emitting source located near to center of our galaxy. In Table 2, we display the estimated values of differences in arrival time corresponding to all three neutrino flavors assuming their masses equal to the experimentally obtained upper bound limits [36] and between photon (i.e., $\Delta T_{n} \vert^{m=0}_m, \left\{n=1,2 \right\}$) in seconds, for the explicit contribution of DM and DE. For $\nu_e$ and $\nu_{\mu}$, we choose the typical value of energy of the observed signal $\varepsilon = 10$ MeV, however for $\nu_{\tau}$, we choose the energy of the observed signal to be $\varepsilon = 100$ MeV, owing to the large upper bound limit of its mass. 

\begin{table}
\large
\centerline{\large Table 2}
\centerline{\large Estimated values of $\Delta T_{n} \vert^{m=0}_m, \left\{n=1,2 \right\}$ for potential astrophysical events} 
\centerline{\large when $\varepsilon =100 $ MeV for $\nu_{\tau}$ and 10 MeV for other neutrino flavors and assuming} 
\centerline{\large $m_{\nu_e} \sim  2 \, {\rm eV} \, , m_{\nu_{\mu}} \sim  0.19 \, {\rm MeV} \, , m_{\nu_{\tau}} \sim  18.2 \, {\rm MeV} $}
\begin{center}
\begin{tabular}{cccccccccccc}
\hline
\hline
\noalign{\vskip 2mm}
${\rm Scenarios}$ & ${\rm G.M}$ & $M/M_{\odot}$ & $r_o $ & $D_S$ &  $\nu$ & $\Delta T_{1} \vert^{m=0}_m$ & $\Delta T_{2} \vert^{m=0}_m$  \\
$ $ & $ $ & $ $ & $ $ & $ $ & $ $ & $\rm (sec)$ & $\rm (sec)$ \\
\hline
\hline 
\noalign{\vskip 2mm}
$R >> r_p$  &  ${\rm Our \, galaxy}$ & $\sim 10^{11}$  & $\sim 10 \, {\rm Kpc}$ & $\sim 50 \, {\rm Kpc}$ & $ \nu_e $ & $\sim 1.1 \times 10^{-8}$ & $\sim 1.6 \times 10^{-13}$\\
\noalign{\vskip 2mm}
$ $  &  $ $ & $ $  & $ $  & $ $ & $ \nu_{\mu} $ & $\sim 103$ & $\sim 1.4 \times 10^{-3}$ \\
\noalign{\vskip 2mm}
$ $  &  $ $ & $ $  & $ $  & $ $ & $ \nu_{\tau} $ & $\sim 9.5 \times 10^{3} $ & $\sim 0.13$\\
\noalign{\vskip 2mm}
$ $  &  ${\rm Local \, group}$  & $\sim 10^{12}$ & $\sim  0.52 \, {\rm Mpc}$ & $\sim  10 \, {\rm Mpc}$ & $\nu_e $ &  $\sim 2.8 \times 10^{-4}$ & $\sim 8.8 \times 10^{-8} $\\
\noalign{\vskip 2mm}
$ $  &  $ $ & $ $  & $ $  & $ $ & $ \nu_{\mu} $ &  $\sim 2.5 \times 10^{6}$  & $\sim 797$ \\
\noalign{\vskip 2mm}
$ $  &  $ $ & $ $  & $ $  & $ $ & $ \nu_{\tau} $ & $\sim 2.3 \times 10^{8}$ & $\sim 7.3 \times 10^{4}$\\ 
\noalign{\vskip 2mm}
$ $  &  ${\rm Local \, supercluster}$  & $\sim 10^{15}$ & $\sim  16 \, {\rm Mpc}$ & $\sim  50 \, {\rm Mpc}$ & $\nu_e $ &  $\sim 45$ & $\sim 4.5 \times 10^{-4} $\\
\noalign{\vskip 2mm}
$ $  &  $ $ & $ $  & $ $  & $ $ & $ \nu_{\mu} $ &  $\sim 4 \times 10^{11}$  & $\sim 4.1 \times 10^{6}$ \\
\noalign{\vskip 2mm}
$ $  &  $ $ & $ $  & $ $  & $ $ & $ \nu_{\tau} $ & $\sim 3.7 \times 10^{13}$ & $\sim 3.7 \times 10^{8}$\\ 
\hline
\noalign{\vskip 2mm}
$R \sim r_p$  &  ${\rm Our \, galaxy}$ & $\sim 10^{11}$  & $\sim 10 \, {\rm Kpc}$ & $\sim 10 \, {\rm Kpc}$ & $ \nu_e $ & $\sim 2.3 \times 10^{-9}$ & $\sim 3.3 \times 10^{-14} $\\
\noalign{\vskip 2mm}
$ $  &  $ $ & $ $  & $ $  & $ $ & $ \nu_{\mu} $ & $\sim 20.6 $ & $\sim 2.9 \times 10^{-4} $\\
\noalign{\vskip 2mm}
$ $  &  $ $ & $ $  & $ $  & $ $ & $ \nu_{\tau} $ & $\sim 1.9 \times 10^{3} $ & $\sim 0.03 $\\
\noalign{\vskip 2mm}
$ $  &  ${\rm Local \, group}$  & $\sim 10^{12}$ & $\sim  0.52 \, {\rm Mpc}$ & $\sim  0.52 \, {\rm Mpc}$ & $\nu_e $ &  $\sim 1.4 \times 10^{-5}$ & $\sim 4.6 \times 10^{-9}$ \\
\noalign{\vskip 2mm}
$ $  &  $ $ & $ $  & $ $  & $ $ & $ \nu_{\mu} $ & $\sim 1.3 \times 10^{5}$ & $\sim 41$ \\
\noalign{\vskip 2mm}
$ $  &  $ $ & $ $  & $ $  & $ $ & $ \nu_{\tau} $ & $\sim 1.2 \times 10^{7}$ & $\sim 3.8 \times 10^{3} $\\
\noalign{\vskip 2mm}
$ $  &  ${\rm Local \, supercluster}$  & $\sim 10^{15}$ & $\sim  16 \, {\rm Mpc}$ & $\sim  16 \, {\rm Mpc}$ & $\nu_e $ &  $\sim 14 $ & $\sim 1.5 \times 10^{-4}$ \\
\noalign{\vskip 2mm}
$ $  &  $ $ & $ $  & $ $  & $ $ & $ \nu_{\mu} $ & $\sim 1.3 \times 10^{11}$ & $\sim 1.4 \times 10^{6}$ \\
\noalign{\vskip 2mm}
$ $  &  $ $ & $ $  & $ $  & $ $ & $ \nu_{\tau} $ & $\sim 1.2 \times 10^{13}$ & $\sim 1.2 \times 10^{8} $\\
\hline
\hline
\end{tabular}
\end{center}
\end{table}

When we consider difference in arrival times of two neutrinos with different energies we should get the time delay of the same order as displayed in Table 2 for difference in arrival times of photon and neutrino unless two energies are very close. Measurement of the differences in arrival times of neutrinos with different energies should provide information about gravitational delay as neutrinos are expected to emit within a second in explosions whereas the time difference between the emission of the neutrinos and the optical brightening at the source is somewhat controversial. The results as displayed in Table 2 suggest that  owing to the small upper bound mass of $\nu_{e}$ measuring gravitational time delay of electron neutrinos caused by DE/DM is only feasible when the gravitating system is local supercluster or even larger systems. The distance scale involve is few tens of Mpc which should be detectable by the low energy extension of ICECUBE [38] and some other upcoming/proposed neutrino telescopes and thereby the results found here is also physically meaningful. For the mentioned scenarios, the ratio of gravitational time delay caused by DM and DE to the Minkowskian time delay are $\beta r_{o}$ and $\Lambda r_{o}^{2}/6$ respectively which numerically equal to $\sim 10^{-7}$ and $\sim 10^{-11}$ when the time delay is caused by our galaxy. If the time delay is caused by the local group, these ratios become $\sim 5 \times 10^{-6}$ and $\sim 5 \times 10^{-10}$ respectively and for local supercluster they take the value $\sim 0.16 $ and $\sim 5 \times 10^{-7}$ respectively. So a good idea about the source distance is needed to discriminate the DE/DM contribution on time delay from Minkowskian and pure gravitational time delay.

\section{Conclusion}

We obtain analytical expression for gravitational time delay of particles with non-zero rest mass in presence of dark energy/matter. We found that a measurement of gravitational time delay involving Mpc distance-scale should detect the contribution of galactic rotation curve description under conformal gravity as well as dark energy in terms of cosmological constant. Hence if such a measurement can be realized in future, it should cross check the validity of the potential linear in radial distance that describes the flat rotation curve of spiral galaxies consistently. The magnitude of cosmological constant also may be verified from such measurements. An interesting observation is that for cosmological constant description of dark energy, the source distance does not appear in the difference of arrival time between particles with same mass but different energies or between particle with mass $m$ and photon, but for some other dark energy models such as the DGP braneworld gravity or massive gravity description the stated difference of arrival times do contain source distance term and hence can be very large. This feature, therefore, can distinguish the alternative dark energy models from cosmological constant, at least in principle. 

An important question is which particle to be used for gravitational time delay measurement. Probing gravity through gravitational time delay effect of high energy neutrons originated outside of our solar system is not feasible owing to their short mean life ($\sim 15$ minutes); even ultra high energy neutrons ($>10^{18} {\rm {eV}}$) with Lorentz boosted lifetime would not travel a distance-scale of not more than $10$ Kpc. Neutrinos seem the only viable candidate for the stated purpose. Being weakly interacting particle, neutrino can provide deeper information about both relic and distant Universe. They are messenger of extreme conditions inside SN cores. Core-collapse SNe both galactic and extragalactic are predictable rich sources of neutrinos. Despite the fact that till date SN 1987A is the only detectable SN source of neutrinos, the present generation neutrino detectors e.g., Icecube neutrino telescope, with certain conditions, might be able to detect SNe beyond $10$ Mpc, while furnishing between $10$ and $41$ regular core-collapse SN detections per decade. Besides, high energy cosmic rays (HECRs) which plausibly originate from active galactic nuclei (AGNs), relativistic extragalactic jets or GRBs located mostly at cosmological distances are other profuse sources of very high energy neutrinos ($>$ Tev range). 

A major problem with neutrino, however, is the uncertainty of their mass as well as its small magnitude. So far the exact mass of any neutrino flavour is not known; experiments provide only the upper bound of its mass. However, electron neutrino mass is expected to pin down by the future neutrino experiments such as JUNO [39], KATRIN [40]. Also the study of cosmology leads to some useful information on the mass scale of light neutrinos [1]. Moreover, the lower bound mass of electron neutrino (few meV) may put some constraint on the dark energy parameters from a high precision gravitational time delay measurements. Conversely, the supernova neutrino observation may put some restriction on the mass of the neutrinos, particularly muon and tau neutrinos. There are two (model independent) approaches of measurement of neutrino mass: time-of-flight measurements and precision investigations of weak decays. Owing to our imprecise knowledge about poin mass, investigation of weak decays put much more looser upper bounds on the mass of muon and tau neutrinos. The study of neutrinos from the supernova SN1987a in the Large Magellanic Cloud, employing the time of flight approach yield upper limits of 5.7 eV (95\% CL) on electron neutrino mass [41]. Future detection of neutrinos from supernova explosions is expected to improve the stated limit as well as to impose new limit on masses of muon and tau neutrinos. Obviously the special relativistic term of equations (14)-(17) will play the dominant role for imposing the mass constraint. The dark matter contribution, being the second largest contributor, also may be important, particularly when the gravitating system is local super-cluster or super-cluster and thereby may set the precision limit of the mass determination.




\appendix
\section{}


Here we furnish the correction to the gravitational time delay due to the coupling terms to the first order corresponding to DM/DE models in the Eqs. (10) - (17) (i,e., the cross term between $M$ and $\beta_{i}$) . As the correction due to cross terms will not affect the total transit time or the gravitational time delay considerably, here we only 
show the expressions (due to algebraic simplicity) considering the source `$S(R)$' to be situated far away from the spherical mass distribution about which the signal suffers gravitational time delay, i.e., $R >> r_p$, the scenario already described in \S IV after Eq. (19). \\

Corresponding to $n = 1,2$ for DM/DE, the total transit time $T_1 \left(r_o, R \right)$ and $T_2 \left(r_o, R \right)$ and their corresponding gravitational time delay will be enhanced considering the coupling terms to the first order by the quantity $\delta t_1 \left (r_o, R \right) \, \vert_{\beta M}$ and $\delta t_2 \left (r_o, R \right) \, \vert_{\Lambda M}$, respectively, given by 
\begin{align}
\delta t_1 \left (r_o, R \right) \, \vert_{\beta M}  \approx \frac{D_S \, GM \beta r_o}{c^2} \, \left[\frac{4}{r_o} \left(1+ 
\frac{m^2}{2\varepsilon^2}  \right) - \frac{\sqrt{D^2_S + r^2_o}+2r_o}{2 \left(\sqrt{D^2_S + r^2_o}+r_o \right)^2 } \right] 
\label{A1}
\end{align}
\begin{align}
\delta t_2 \left (r_o, R \right) \, \vert_{\Lambda M} \approx \frac{D_S \, GM \Lambda}{6c^2} \, \left[4 \sqrt{D^2_S + r^2_o} - 2r_o \frac{m^2}{\varepsilon^2} \right]
\label{A2}
\end{align}

Accordingly in Eqs. (14) and (15), corresponding to $n=1$, for DM, the correction term due to coupling to the first order will reduce the corresponding gravitational time delay by quantities $ \sim \frac{2 \, D_S \, G M \beta \, m^2}{c^2 \, \varepsilon^2}$ and $ \sim \frac{2 \, D_S \, G M \beta \, m^2}{c^2} \left( \frac{1}{\varepsilon^2_1} - \frac{1}{\varepsilon^2_2} \right) $, using the condition described above. Similarly, corresponding to $n=2$, for DE, the correction term due to coupling to the first order will also reduce the corresponding gravitational time delay by quantities $ \sim \frac{D_S \, G M  \lambda \, r_o \, m^2}{3 \, c^2 \,  \varepsilon^2}$ and $ \sim \frac{D_S \, G M \lambda \, r_o \, m^2}{3 \, c^2} \left( \frac{1}{\varepsilon^2_1} - \frac{1}{\varepsilon^2_2} \right) $, accordingly in 
Eqs. (16) and (17), respectively.

\end{document}